\documentclass[11pt]{article}
\usepackage{amsmath,amsfonts,latexsym,graphicx,amssymb}
\usepackage{amsfonts}
\usepackage{bm}

\newcommand{\id}{\mbox{id}}
\newcommand{\ot}{{\,\otimes\,}}
\newcommand{{\Cd}}{{\mathbb{C}^d}}
\newcommand{{\C}}{{\mathbb{C}}}
\newtheorem{theorem}{Theorem}
\newtheorem{definition}{Definition}

\def\<{\langle}
\def\>{\rangle}
\date{}

\begin{document}
\title{\textbf{On partially entanglement breaking channels}}
\author{Dariusz Chru\'sci\'nski and Andrzej Kossakowski \\
Institute of Physics, Nicolaus Copernicus University,\\
Grudzi\c{a}dzka 5/7, 87--100 Toru\'n, Poland}

\maketitle

\begin{abstract}
Using well known duality between quantum maps and states of
composite systems we introduce the notion of Schmidt number of a
quantum channel. It enables one to define  classes of quantum
channels which partially break quantum entanglement. These classes
generalize the well known class of entanglement breaking channels.
\end{abstract}

\section{Introduction}

In quantum information theory \cite{Nielsen} a quantum channel is
represented by a completely positive trace preserving map (CPT)
between states of two quantum systems living in ${\cal H}_A$ and
${\cal H}_B$. Consider ${\cal H}_A = {\cal H}_B = \Cd$. Then the
states of both systems are defined by semi-positive elements from
$M_d \cong \mathbb{C}^d \otimes \mathbb{C}^d$. Due to the
Kraus-Choi representation theorem \cite{Kraus} any CPT map
\begin{equation}\label{CPT}
    \Phi\ :\ M_d  \ \longrightarrow\ M_d \
    ,
\end{equation}
may be represented by
\begin{equation}\label{KK}
    \Phi(\rho) = \sum_\alpha\, K_\alpha\, \rho\, K^*_\alpha\ ,
\end{equation}
where the Kraus operators $K_\alpha \in M_d$ satisfies
trace-preserving condition $\sum_\alpha\, K_\alpha^*\, K_\alpha =
I_d$. It is, therefore, clear that all the properties of $\Phi$
are encoded into the family $K_\alpha$. In the present paper we
show how the structure of $\Phi$ depends upon the rank of Kraus
operators. In particular it is well known
\cite{Shor-Horodecki,Ruskai} that if all $K_\alpha$ are rank one
then $\Phi$ defines so called entanglement breaking channel (EBT),
that is, for any state $\rho$ from $M_d \ot M_d$, $ (\id_d \otimes
\Phi)\rho\ $
 is separable in $M_d \ot M_d$.

\begin{definition} We call a channel (\ref{CPT}) an
 $r$--partially entanglement breaking channel ($r$--PEBT) iff for an arbitrary
 $\rho$
\begin{equation}\label{}
    \mbox{SN}[(\id_d \otimes \Phi)\rho] \leq r \ ,
\end{equation}
where $\mbox{SN}(\sigma)$ denotes the Schmidt number of $\sigma$.
\end{definition}

Clearly, EBT channels are 1--PEBT.  Let us recall
\cite{Terhal-Horodecki} that
\begin{equation}\label{SN-rho}
    \mbox{SN}(\sigma) = \min_{p_k,\psi_k}\, \left\{ \,
    \max_{k}\, \mbox{SR}(\psi_k)\, \right\}  \ ,
\end{equation}
where the minimum is taken over all possible pure states
decompositions
\begin{equation}\label{}
    \sigma = \sum_k \, p_k\, |\psi_k\>\<\psi_k|\ , \nonumber
\end{equation}
with $p_k\geq 0$, $\sum_k\, p_k =1$ and $\psi_k$ are normalized
vectors in $\mathbb{C}^d \otimes \mathbb{C}^d$. The Schmidt rank
SR$(\psi)$ denotes the number of non-vanishing Schmidt
coefficients in the Schmidt decomposition of $\psi$. This number
characterizes the minimum Schmidt rank of the pure states that are
needed to construct such density matrix. It is evident that $ 1
\leq \mbox{SN}(\rho) \leq d $ and $\rho$ is separable iff
$\mbox{SN}(\rho) =1 $. Moreover,  it was proved
\cite{Terhal-Horodecki} that the Schmidt number is non-increasing
under local operations and classical communication.

 Let us denote by $S_k$ the set of density matrices on
$\mathbb{C}^d \otimes \mathbb{C}^d$ that have Schmidt number at
most $k$. One has ${\cal S} = S_1 \subset S_2 \subset \ldots
\subset S_d = {\cal P}$ with ${\cal S}$ and ${\cal P}$ being the
sets of separable and all density matrices, respectively. Recall,
that a positive map $ \Lambda : M_d \longrightarrow M_d$ is
$k$-positive, if $(\id_k \otimes \Lambda)$ is positive on
$M_k\otimes M_d$. Due to Choi \cite{Choi} $\Lambda$ is completely
positive iff it is $d$-positive. Now, $\Lambda$ is $k$-positive
iff $(\id_d \otimes \Lambda)$ is positive on  $S_k$. The set of
$k$-positive maps which are not $(k+1)$-positive may be used to
construct  a Schmidt number witness operator $W$ which is
non-negative on all states in $S_{k-1}$, but detects at least one
state $\rho$ belonging to $S_k$ \cite{Bruss1,Bruss2} (see also
\cite{Eisert}), i.e.
\begin{equation}\label{}
\mbox{Tr}\, (W\sigma) \geq 0 \ , \ \ \ \ \sigma \in S_{k-1} \ ,
\end{equation}
and there is a $\rho \in S_k$ such that $\mbox{Tr}\, (W\rho) < 0$.

In the next section we investigate basic properties of PEBT
channels. Then in section  \ref{Multi} we generalize the
discussion to multipartite entangled states.

\section{Properties of PEBT channels}

Using well know duality between quantum CPT maps (\ref{CPT}) and
states of the composite quantum system living in $\mathbb{C}^d
\otimes \mathbb{C}^d$ \cite{Zyczkowski,Kossakowski} we may assign
a Schmidt number to any CPT map. Take any CPT map $\Phi$ and
define a state \cite{Jam}
\begin{equation}\label{J}
    \rho_\Phi = (\id_d \otimes \Phi)\, P^+_d  \ ,
\end{equation}
where $P^+_d = |\psi^+_d\>\<\psi^+_d|$ with $\psi^+_d =
d^{-1/2}\sum_k\, e_k \otimes e_k$ being a maximally entangled
state in $\mathbb{C}^d \otimes \mathbb{C}^d$ ($e_k\, ; \
k=1,2,\ldots,d$ denote the orthonormal base in $\mathbb{C}^d$).

\begin{definition}
A Schmidt number of $\Phi$ is defined by
\begin{equation}\label{}
    \mbox{SN}(\Phi) = \mbox{SN}(\rho_\Phi)\ ,
\end{equation}
where $\rho_\Phi$ stands for the `dual' state defined in
(\ref{J}).
\end{definition}

Actually, in \cite{Kossakowski} a CPT map $\Phi : M_d
\longrightarrow M_d$ was called an $r$--CPT iff SN$(\Phi) \leq r$.
We show that $r$--PEBT channels are represented by $r$--CPT maps.

Note, that using Kraus decomposition (\ref{KK}) we may express the
Schmidt number of $\Phi$ in analogy to (\ref{SN-rho}) as follows:
\begin{equation}\label{SN-Phi}
    \mbox{SN}(\Phi) = \min_{K_\alpha}\, \left\{ \,
    \max_{\alpha}\, \mbox{rank}\, K_\alpha\, \right\}  \ .
\end{equation}
 The analogy between
(\ref{SN-rho}) and (\ref{SN-Phi}) is even more visible if we make
the following observation: any vector $\psi \in \mathbb{C}^d
\otimes \mathbb{C}^d$ may be written as $\psi = \sum_{i,j=1}^d
x_{ij} e_i \otimes e_j$ and hence, introducing a $\psi$-dependent
operator $F \in M_d$ such that $x_{ij}= \<j|F|i\>  $, one has
\begin{equation}\label{psi-F}
    \psi = \sum_{i=1}^d\, e_i \otimes F e_i \ .
\end{equation}
Using the maximally entangled state $\psi^+_d$ it may be rewritten
in perfect analogy to (\ref{J}):
\begin{equation}\label{J-F}
    \psi = \sqrt{d}\, (\id_d \ot F)\psi^+_d\ .
\end{equation}
 Clearly, the above formula
realizes an isomorphism between $\mathbb{C}^d \otimes
\mathbb{C}^d$ and $M_d$. Note, that the normalization condition
$\<\psi|\psi\> = 1$ implies $\mbox{Tr}(F^* F)=1$. Moreover, two
vectors $\psi_1$ and $\psi_2$ are orthogonal iff the corresponding
operators $F_1$ and $F_2$ are trace-orthogonal, i.e.
$\mbox{Tr}(F_1^\dag F_2)=0$. It is evident that $\mbox{SR}(\psi) =
\mbox{rank}\, F$. Moreover, the singular values of $F$ are nothing
but the Schmidt coefficients of $\psi$. Hence, the separable pure
states from $\mathbb{C}^d \otimes \mathbb{C}^d$ correspond to rank
one operators from $M_d$.

Consider now the  corresponding one-dimensional projector
$|\psi\>\<\psi|$. It  may be written as
\begin{equation}\label{psi-FF}
    |\psi\>\<\psi| = \sum_{i,j=1}^d\, e_{ij} \ot  F e_{ij}  F^* \
    ,
\end{equation}
with  $\mbox{Tr}(F^\dag F)=1$. In (\ref{psi-FF}) a rank one
operator $e_{ij} \in M_d$ equals to $|i\>\<j|$ in Dirac notation.
Hence the Schmidt class $S_k$ may be defined as follows: $\rho \in
S_k$ iff
\begin{equation}\label{}
    \rho = \sum_\alpha\, p_\alpha P_\alpha\ ,
\end{equation}
with $p_\alpha\geq 0$, $\sum_\alpha\, p_\alpha=1$ and
\begin{equation}\label{P-alpha}
    P_\alpha =  \sum_{i,j=1}^d\, e_{ij} \ot F_\alpha e_{ij}
    F_\alpha^*\ ,
\end{equation}
with $\mbox{rank}\, F_\alpha \leq k$,  and $\mbox{Tr}(F_\alpha
F^*_\alpha)=1$. That is, $S_k$ is a convex combination of one
dimensional projectors corresponding to $F$'s of rank at most $k$.

\begin{theorem} A quantum channel $\Phi \in $ $r$--PEBT iff
$\ SN(\Phi)\leq r$.
\end{theorem}

{\it Proof.} Note, that $\mbox{SN}(\Phi) \leq r$  iff there exists
a Kraus decomposition such that all Kraus operators $K_\alpha$
satisfy $\mbox{rank}\,K_\alpha \leq r$. Indeed, using (\ref{KK})
and (\ref{P-alpha}) one has
\begin{eqnarray}\label{}
(\id_d \ot \Phi) \, P^+_d = \sum_{i,j=1}^d \, e_{ij} \ot
\Phi(e_{ij})  = \sum_\alpha\, p_\alpha P_\alpha\ , \nonumber
\end{eqnarray}
with
\[ p_\alpha = \frac 1d \mbox{Tr}(K^\dag_\alpha K_\alpha)\ , \ \ \
\ \  F_\alpha = \frac{1}{\sqrt{dp_\alpha}}\,K_\alpha \ . \] The
above relations simply translate the isomorphism between states
and CPT maps in terms of operators $K_\alpha$ and $F_\alpha$.
Suppose now that $\Phi$ is $r$-PEBT and let $\rho$ be an arbitrary
state in $M_d$
\[ \rho = \sum_\beta p_\beta\, \sum_{i,j=1}^d\, e_{ij} \ot F_\beta\, e_{ij}\, F_\beta^*\ , \]
with arbitrary $F_\alpha \in M_d$ such that $\mbox{Tr}(F_\alpha
F^*_\alpha)=1$.  One has
\begin{eqnarray}\label{}
 (\id_d \otimes \Phi)\rho  \ = \ \sum_{\alpha,\beta}\,p_{\alpha\beta}\,
\sum_{i,j=1}^d\,  e_{ij} \otimes \widetilde{F}_{\alpha\beta}
e_{ij} \widetilde{F}_{\alpha\beta}^*\ ,
\end{eqnarray}
with
\[  p_{\alpha\beta} \ =\ \frac 1d\,
\mbox{Tr}(K_\alpha K^*_\alpha)\, p_\beta\ , \ \ \ \
\widetilde{F}_{\alpha\beta} \ = \
\sqrt{\frac{dp_\beta}{p_{\alpha\beta}}}\, K_\alpha F_\beta\ ,
\]
where $K_\alpha$ are Kraus operators representing an $r$--CPT map
$\Phi$ satisfying rank$K_\alpha\leq r$. Now,
\begin{equation}\label{}
    \mbox{rank}\, (K_\alpha F_\beta) \leq \min \{\mbox{rank}\,
    K_\alpha\, ,
    \mbox{rank}\,F_\beta\} \leq r \ , \nonumber
\end{equation}
and hence $(\id_d \ot \Phi)\, \rho \in S_r$. The converse follows
immediately. \hfill $\Box$

\noindent As a corollary note that since $\mbox{rank}\, (K_\alpha
F_\beta) \leq \mbox{rank}\, F_\beta$ one finds
\begin{equation}\label{}
    \mbox{SN}( (\id_d \ot \Phi)\, \rho ) \leq \mbox{SN}(\rho) \
    ,
\end{equation}
which shows that indeed SN does not increase under a local
operation defined by $\id_d \ot \Phi$.

\begin{theorem}A map $\Phi$ is $r$-CPT iff $\Lambda \circ \Phi$
is CPT for any $r$-positive map $\Lambda$.
\end{theorem}

 {\it Proof.} Suppose that $\Phi$ is $r$-CPT and take an
arbitrary $k$-positive $\Lambda$:
\begin{equation}\label{Proof}
    (\id_d \ot \Lambda \circ \Phi) \, P^+_d = (\id_d \ot \Lambda)\left[ (\id_d \ot \Phi) \,
    P^+_d\right] \geq 0 \ , \nonumber
\end{equation}
since $(\id_d \ot \Phi) \, P^+_d \in S_r$. Conversely, let
$\Lambda \circ \Phi$ be CPT for any $r$-positive  $\Lambda$, then
$ (\id_d \ot \Lambda \circ \Phi) \, P^+_d \geq 0 $ implies that
$(\id_d \ot \Phi) \,P^+_d \in S_r$ and hence $\Phi$ is $r$-CPT.
Actually, the same is true for $\Phi \circ \Lambda$. \hfill $\Box$

To introduce another class of quantum operations let us recall the
notion of co-positivity: a map $\Lambda$ is $r$--co-positive iff
$\tau \circ \Lambda$ is $r$-positive, where $\tau$ denotes
transposition in $M_d$. In the same way $\Phi$ is completely
co-positive (CcP) iff $\tau \circ \Phi$ is CP. Let us define the
following convex subsets in $M_d \otimes M_d$: $S^r = (\id_d \ot
\tau)\, S_r$. One obviously has: $S^1 \subset S^2 \subset \ldots
\subset S^n$. Note, that $S^1=S_1 ={\cal S}$ and $S_n \cap S^n$ is
a set of all PPT states.

Now, following \cite{Kossakowski} we call a CcPT map $\Phi$ an
$(r,s)$-CPT if
\begin{equation}\label{r-CPT}
(\id_d \ot \Phi) \, P^+_d \in S_r \cap S^s\ ,
\end{equation}
that is
\[  \rho_\Phi \in S_r \ \  \ \ \mbox{and}\ \ \ \
(\id_d \ot \tau)\rho_\Phi \in S_s \ . \]
 Hence, if $\rho_\phi$ is
a PPT state, then $\Phi$ is $(r,s)$-CPT for some $r$ and $s$. In
general there is no relation between $(r,s)$-CPT and $(k,l)$-CPT
for arbitrary $r,s$ and $k,l$. However, one has
\[   (1,1)\mbox{-CPT} \subset (2,2)\mbox{-CPT} \subset \ldots \subset
(n,n)\mbox{-CPT} \ , \] and $(n,n)\mbox{-CPT} \equiv \mbox{CPT}
\cap \mbox{CcPT}$.

\noindent {\bf Theorem 3:} A map $\Phi$ is $(r,s)$-CPT iff for any
$r$-positive map $\Lambda_1$ and $s$--co-positive map $\Lambda_2$
the composite map $\Lambda_1 \circ \Lambda_2 \circ\Phi$ is CPT.

\section{Examples}

\noindent {\bf Example 1:} Let us consider so called isotropic
state in $d$ dimensions
\begin{equation}\label{}
{\cal I}_\lambda = \frac{1-\lambda}{d^2} I_d \otimes I_d + \lambda
P^+_d \ ,
\end{equation}
with $-1/(d^2-1) \leq \lambda \leq 1$. It is well known
\cite{Horodecki} that ${\cal I}_\lambda$ is separable iff $\lambda
\leq 1/(d+1)$. Now, let $\Psi : M_d \longrightarrow M_d$ be an
arbitrary positive trace preserving map and define a CPT map
$\Phi_\lambda$ by
\begin{equation}\label{}
    (\id_d \ot \Phi_\lambda) P^+_d = (\id_d \ot \Psi){\cal I}_\lambda\ .
\end{equation}
One easily finds
\begin{equation}\label{}
\Phi_\lambda(\rho) =  \frac{1-\lambda}{d}\,\mbox{Tr}\rho\,  I_d +
\lambda\Psi(\rho)\ .
\end{equation}
Clearly, for $\lambda \leq 1/(d+1)$ (i.e. when ${\cal I}_\lambda$
is separable) $\Phi_\lambda$ is $(1,1)$-CPT, i.e. both
$\Phi_\lambda$ and $\tau \circ \Phi_\lambda$ are EBT.

\noindent {\bf Example 2:} Let us rewrite an isotropic state
${\cal I}_\lambda$ in terms of  fidelity $f=\mbox{Tr}({\cal
I}_\lambda\, P^+_d)$:
\begin{equation}\label{}
    I_f = \frac{1-f}{d^2-1} (I_d \ot I_d - P^+_d) + fP^+_d \ .
\end{equation}
It was shown in \cite{Terhal-Horodecki} that SN$({\cal I}_f) = k$
iff
\begin{equation}\label{f}
 \frac{k-1}{d} < f \leq \frac{k}{d} \ .
\end{equation}
Defining  a CPT map $\Phi_f$
\begin{equation}\label{}
    (\id_d \ot \Phi_f) P^+_d = {\cal I}_f \ ,
\end{equation}
one finds
\begin{equation}\label{}
    \Phi_f(\rho) = \frac{1-f}{d^2-1}\, \mbox{Tr}\rho\, I_d +
    \frac{d^2f-1}{d^2-1}\, \rho\ .
\end{equation}
This map is $k$--CPT iff $f$ satisfies (\ref{f}) and hence it
represents an $r$--PEBT channel.

\noindent {\bf Example 3:}  Consider
\begin{equation}\label{FU}
    \rho = \sum_{\alpha=1}^{d^2}\, p_\alpha\,
    \sum_{i,j=1}^d\, e_{ij} \ot F_\alpha\, e_{ij}\, F^*_\alpha\ ,
\end{equation}
where
\begin{equation}\label{}
 p_\alpha\geq 0\ ,\ \ \ \  \ \ \
\sum_{\alpha=1}^{d^2}\, p_\alpha=1\ ,\ \ \ \ \ \ \
 F_\alpha = \frac{U_\alpha}{\sqrt{d}}\ ,
\end{equation}
and $U_\alpha$ defines a family of unitary operators from $U(d)$
such that
\begin{equation}\label{}
  \mbox{Tr}(U_\alpha\, U^*_\beta) = \delta_{\alpha\beta} \ , \ \
\ \ \ \ \alpha,\beta = 1,2,\ldots,d^2\ .
\end{equation}
The corresponding `dual' quantum channel $\Phi$ is given by
\begin{equation}\label{Phi-d}
    \Phi(\sigma) = \sum_{\alpha=1}^{d^2}\, K_\alpha\, \sigma\,
    K^*_\alpha\ ,
\end{equation}
with $K_\alpha= \sqrt{p_\alpha}\, U_\alpha$. Note, that for
$p_\alpha = 1/d^2$ one obtains a completely depolarizing channel,
i.e.
\begin{equation}\label{depolar}
    \frac{1}{d^2}\,  \sum_{\alpha=1}^{d^2}\, U_\alpha\, e_{ij}\,
    U^*_\alpha\ = \delta_{ij}\ .
\end{equation}
Now, following \cite{Tomiyama} consider a map
\begin{equation}\label{}
    \Lambda_\mu(\sigma) = I_d\, \mbox{Tr}\, \sigma - \mu \sigma\ ,
\end{equation}
which is $k$ (but not $(k+1)$)--positive for
\begin{equation}\label{k-mu}
    \frac{1}{k+1} \leq \mu \leq \frac 1k\ .
\end{equation}
One has
\begin{eqnarray}\label{}
    (\id_d \ot \Lambda_\mu) \rho &=& \sum_{\alpha=1}^{d^2}\, p_\alpha\,
    \sum_{i,j=1}^d\, e_{ij} \ot \left[ I_d\, \mbox{Tr}(F_\alpha\, e_{ij}\,
    F^*_\alpha) - \mu\, F_\alpha\, e_{ij}\,
    F^*_\alpha \right] \nonumber \\
    &=& \frac 1d\, I_d \ot I_d - \sum_{\alpha=1}^{d^2}\, \mu p_\alpha\,
    \sum_{i,j=1}^d\, e_{ij} \ot F_\alpha\, e_{ij}\, F^*_\alpha
    \nonumber \\
    &=& \frac 1d\, \sum_{\alpha=1}^{d^2}\, (1 - d\mu p_\alpha)\,
    \sum_{i,j=1}^d\, e_{ij} \ot F_\alpha\, e_{ij}\, F^*_\alpha\ ,
\end{eqnarray}
where we have used (\ref{depolar}). It is therefore clear that if
for some $1\leq \alpha \leq d^2$, $\, p_\alpha > 1/(d\mu)$ and
$\mu$ satisfies (\ref{k-mu}), then $\mbox{SN}(\rho) \geq k+1$.
Equivalently, a `dual' quantum channel (\ref{Phi-d}) belongs to
$\{\, d$--PEBT $-$ $k$--PEBT$\}$.

\section{PEBT channels and multipartite entanglement}
\label{Multi}

Consider now a multipartite entangled state living in ${\cal H} =
( \mathbb{C}^d)^{\ot N}$ for some $N\geq 2$. Any $\psi \in {\cal
H}$ may be written as follows:
\begin{equation}\label{multi-F}
    \psi = \sum_{i_1,\ldots,i_K=1}^d \, e_{i_1} \ot \ldots \ot e_{i_K}
    \ot F ( e_{i_1} \ot \ldots \ot e_{i_K}) \ ,
\end{equation}
where $F$ is an operator
\[  F \ :\  (\mathbb{C}^d)^{\ot K}\ \longrightarrow\ (\mathbb{C}^d)^{\ot
N-K}  \ , \] and $1\leq K \leq N-1$. Again, normalization of
$\psi$ implies $\mbox{Tr}(F^*F)=1$. Clearly, such representation
of $\psi$ is highly non-unique. One may freely choose $K$ and take
$K$ copies of $\Cd$ out of $(\Cd)^{\ot N}$. Any specific choice of
representation depends merely on a specific question we would like
to ask. For example (\ref{multi-F}) gives rise to the following
reduced density matrices:
\begin{equation}\label{}
    \rho_B = \mbox{Tr}_A \, |\psi\>\<\psi| = \mbox{Tr}_{12\ldots
    K}\, |\psi\>\<\psi| \, = \, FF^* \, \in \, M_d^{\ot N-K} \ ,
\end{equation}
and
\begin{equation}\label{}
    \rho_A = \mbox{Tr}_B \, |\psi\>\<\psi| = \mbox{Tr}_{K+1\ldots
    N}\, |\psi\>\<\psi| \, = \, F^*F \, \in \, M_d^{\ot K} \ .
\end{equation}
A slightly different way to represent $\psi$ reads as follows
\begin{equation}\label{}
    \psi = \sum_{i_1,\ldots,i_{N-1}=1}^d \, e_{i_1} \ot \ldots \ot
    e_{i_{N-2}} \ot e_{i_{N-1}}
    \ot F_{i_1\ldots i_{N-2}} e_{i_{N-1}} \ ,
\end{equation}
where
\[  F_{i_1\ldots i_{N-2}} \ :\  \mathbb{C}^d\ \longrightarrow\
\mathbb{C}^d\ ,  \] for any $i_1,\ldots,i_{N-2}=1,2,\ldots,d$.
Now, normalization of $\psi$ implies
\begin{equation}\label{}
\sum_{i_1,\ldots,i_{N-2}=1}^d\, \mbox{Tr}\left( F_{i_1\ldots
i_{N-2}}^*F_{i_1\ldots i_{N-2}}\right) =1 \ .
\end{equation}
 One has the following relation between different
representations:
\begin{equation}\label{}
 \< e_{i_N} |F_{i_1\ldots i_{N-2}}|e_{i_{N-1}}  \>\, =\,
\< e_{i_1} \ot \ldots \ot e_{i_{N-1}}|F|e_{i_N}\> \ .
\end{equation}

\noindent {\bf Example 4.}
 For $N=3$ we have basically three representations:
\begin{equation}\label{3-I}
    \psi = \sum_{i=1}^d\, e_i \ot F e_i \ ,
\end{equation}
\begin{equation}\label{3-II}
    \psi = \sum_{i,j=1}^d \, e_i \ot e_j \ot F'(e_i \ot e_j) \ ,
\end{equation}
 and
\begin{equation}\label{3-III}
    \psi = \sum_{i,j=1}^d \, e_i \ot e_j \ot F_{i}\, e_j  \ ,
\end{equation}
with
\[  F \ :\  \mathbb{C}^d\ \longrightarrow\
(\mathbb{C}^d)^{\ot 2}\ , \ \ \ \ F' = F^T \ :\
(\mathbb{C}^d)^{\ot 2}\ \longrightarrow\ \mathbb{C}^d \ , \ \ \ \
F_{i} \ :\  \mathbb{C}^d\ \longrightarrow\ \mathbb{C}^d \ . \] As
an example take $d=2$ and let us consider two well known 3-qubit
states \cite{GHZ}:
\begin{equation}\label{GHZ}
    |\mbox{GHZ}\> = \frac{1}{\sqrt{2}}\, \left( |000\> + |111\> \right)\ ,
\end{equation}
and
\begin{equation}\label{W}
    |W\> = \frac{1}{\sqrt{3}}\, \left( |100\> + |010\> + |001\> \right)\
    .
\end{equation}
One finds for GHZ--state:
\begin{equation}\label{}
F' = (F_1,F_2) = \frac{1}{\sqrt{2}}\, \left( \begin{array}{cccc}
1&0&0&0\\ 0&0&0&1
\end{array} \right) = F^T  \ ,
\end{equation}
and for W--state:
\begin{equation}\label{}
 \widetilde{F}' = (\widetilde{F}_1,\widetilde{F}_2) = \frac{1}{\sqrt{3}}\,
\left( \begin{array}{cccc} 0&1&1&0\\ 1&0&0&0
\end{array} \right) = \widetilde{F}^T \ .
\end{equation}
Note, that for both states $\mbox{rank}(F) =
\mbox{rank}(\widetilde{F}) =2$. There is, however, crucial
difference between $F_i$ and $ \widetilde{F}_i$:
$\mbox{rank}(F_i)=1$, whereas $\mbox{rank}( \widetilde{F}_1)=2$.
Both states possess genuine 3--qubit entanglement. The difference
consists in the fact that GHZ--state is 2--qubit separable whereas
W--state is 2--qubit entangled \cite{Maciek}:
\begin{equation}\label{}
 \rho^{\mbox{\scriptsize GHZ}}_{\, 23} = \mbox{Tr}_1
|\mbox{GHZ}\>\<\mbox{GHZ}| = \sum_{k=0}^1\, \sum_{i,j=0}^1\,
e_{ij} \ot F_k\, e_{ij}\, F^*_k \ ,
\end{equation}
 with $ \mbox{SN}( \, \rho^{\mbox{\scriptsize GHZ}}_{\, 23}\, ) =
1\, $ ,

and
\begin{equation}\label{}
 \rho^{\mbox{\scriptsize W}}_{\, 23} = \mbox{Tr}_1
|\mbox{W}\>\<\mbox{W}| = \sum_{k=0}^1\, \sum_{i,j=0}^1\, e_{ij}
\ot \widetilde{F}_k\, e_{ij}\, \widetilde{F}^*_k \ ,
\end{equation}
 with $\mbox{SN}( \, \rho^{\mbox{\scriptsize W}}_{\, 23}\, ) = 2\, $.

If $N=2K$ any state vector $\psi \in (\Cd)^{\ot N} = (\Cd)^{\ot K}
\ot (\Cd)^{\ot K}$ may be represented by (\ref{multi-F}) with
\begin{equation}\label{}
 F \ :\  (\mathbb{C}^d)^{\ot K}\ \longrightarrow\ (\mathbb{C}^d)^{\ot
K}  \ .
\end{equation}
Hence, an arbitrary state $\rho$ from $M_d^{\ot K} \ot M_d^{\ot
K}$ reads as follows
\begin{equation}\label{2K-F}
    \rho = \sum_\alpha\, p_\alpha \sum_{i_1,\ldots,i_K=1}^d\,
    \sum_{j_1,\ldots,j_K=1}^d \, e_{i_1j_1} \ot \ldots \ot e_{i_Kj_K}
    \ot F_\alpha ( e_{i_1j_1} \ot \ldots \ot e_{i_Kj_K}) F_\alpha^* \
    .
\end{equation}
Clearly, $\mbox{SN}(\rho) \leq r$ iff $\mbox{rank}(F_\alpha)\leq
r$ for all $F_\alpha$ appearing in (\ref{2K-F}). Then the
corresponding quantum channel
\begin{equation}\label{}
    \Phi \ :\ M_d^{\ot K} \ \longrightarrow\ M_d^{\ot K} \ ,
\end{equation}
possesses Kraus decomposition with $K_\alpha =
\sqrt{d^Kp_\alpha}\, F_\alpha$ and hence is $r$--PEBT. For other
aspects of multipartite entanglement se e.g. \cite{multi}.

\section*{Acknowledgments}

This work was partially supported by the Polish State Committee
for Scientific Research Grant {\em Informatyka i in\.zynieria
kwantowa} No PBZ-Min-008/P03/03.

\end{document}